# Combining chromosomal arm status and significantly aberrant genomic locations reveals new cancer subtypes

Tal Shay[1], Wanyu L. Lambiv[2], Anat Reiner[1,3], Monika E. Hegi[2,4], Eytan Domany[1]

[1]Department of Physics of Complex Systems, Weizmann Institute of Science, Rehovot, Israel;

[2]Laboratory of Tumor Biology and Genetics, Neurosurgery, University Hospital Lausanne (CHUV), Lausanne, Switzerland;

[3]Department of Statistics, University of Haifa, Haifa, Israel;

[4]Institut Suisse de Recherche Experimentale sur le Cancer (ISREC), Epalinges, Switzerland;

# Abstract

Many types of tumors exhibit characteristic chromosomal losses or gains, as well as local amplifications and deletions. Within any given tumor type, sample specific amplifications and deletions are also observed. Typically, a region that is aberrant in more tumors, or whose copy number change is stronger, would be considered as a more promising candidate to be biologically relevant to cancer. We sought for an intuitive method to define such aberrations and prioritize them. We define V, the "volume" associated with an aberration, as the product of three factors: (a) fraction of patients with the aberration, (b) the aberration's length and (c) its amplitude. Our algorithm compares the values of V derived from the real data to a null distribution obtained by permutations, and yields the statistical significance (p-value) of the measured value of V. We detected genetic locations that were significantly aberrant, and combine them with chromosomal arm status (gain/loss) to create a succinct fingerprint of the tumor genome. This genomic fingerprint is used to visualize the tumors, highlighting events that are co-occurring or mutually exclusive. We apply the method on three different public array CGH datasets of Medulloblastoma and Neuroblastoma, and demonstrate its ability to detect chromosomal regions that were known to be altered in the tested cancer types, as well as to suggest new genomic locations to be tested. We identified a potential new subtype of Medulloblastoma, which is analogous to Neuroblastoma type 1.

# Background

**Cancer is characterized by DNA copy number aberrations**

Genes from all bands of the human chromosomes are involved in some commonly occurring tumor associated aberrations (Mitelman et al. 1997). Each solid tumor type displays one of several characteristic combinations of chromosomal gains and losses. There is considerable overlap between the imbalance profiles of the different tumor types, and typically there are more losses than gains (Mertens et al. 1997). It has been shown that in several cancers local DNA copy number aberrations are predictive of outcome (Seeger et al. 1985;Slamon et al. 1987;Kyomoto et al. 1997) or of treatment response (Palmberg et al. 2000;Slamon et al. 2001;Ishiguro et al. 2003). Oncogene activations can result from chromosomal translocations and from gene amplifications. Tumor-suppressor genes' inactivation arises from several mechanisms, including deletions or insertions of various sizes (Vogelstein and Kinzler 2004).

Analysis and interpretation of local aberrations that contribute to cancer development are hindered by the fact that in cancer cells there is loss and gain of whole chromosomes, that may be the cause of the cancer or a by-product of it (cf Marx 2002;Duesberg et al. 2006). While many cancers display karyotypic changes, oncogenic transformation can occur with no chromosomal instability, both in-vitro (Zimonjic et al. 2001) and in-vivo (Lamlum et al. 2000).

**Array CGH as a tool to measure DNA copy number aberrations**

Array Comparative Genomic Hybridization (aCGH) (Solinas-Toldo et al. 1997;Pinkel et al. 1998) is a procedure that provides genome-wide DNA copy number measurement along genomes of mammalian complexity. A control sample and a test sample are competitively hybridized to an array with genomic targets. If the control is diploid, a higher signal of the test sample is indicative of amplification, and a higher control signal indicates deletion. Single-copy decreases and increases from diploid are reliably detected (Pinkel et al. 1998). Several types of genomic targets can be printed on the array. For example, Bacterial Artificial Chromosomes (BACs) are fairly widely used: these markers have a typical length of 150KB, and about 2000-8000 BACS are used to provide coverage of the full human genome. In addition, cDNA probes are also used (Pollack et al. 1999) as well as oligonucleotides (Lucito et al. 2003;Barrett et al. 2004).

**Existing methods for analyzing array CGH data**

Most methods for analysis of aCGH data focus on assigning copy number or status (gain, normal, loss) to every genomic location in single samples (Hodgson et al. 2001;Pollack et al. 2002;Olshen et al. 2004;Lai et al. 2005;Diaz-Uriarte and Rueda 2007;Venkatraman and Olshen 2007). Several such methods were compared (Lai et al. 2005;Willenbrock and Fridlyand 2005), with the conclusion that most algorithms do well in detecting the existence and the width of aberrations for large changes and high signal-to-noise ratio. None of the algorithms, however, detected reliably aberrations with small width and low signal-to-noise ratio. Most studies recognize those aberrations that pass a certain threshold of frequency of appearance or amplitude. In nearly all studies, the selection criteria were either not specified, or set in an arbitrary way (Hodgson et al. 2001;Pollack et al. 2002;Ferreira et al. 2007;Lassmann et al. 2007;Lo et al. 2007).

Considerable effort has been devoted to identify significant and meaningful aberrations, using simultaneously data from multiple samples. Hidden Markov Models, often used to define single sample status, were extended to multiple samples (Shah et al. 2007). Rouveirol et al. (2006) defined recurrent minimal genomic alterations, and incorporated external constraints, such as a range or frequencies of occurrence and a range of signal magnitudes, to filter the observed alterations. Snijders et al. (2005) used aCGH to define minimal common amplified regions and then expression analysis to identify candidate driver genes in amplicons. Diskin et al. (2006) presented a method for testing the significance of aberrations across multiple samples. Their input is a list of aberrations in each sample. They calculate a frequency statistic and a footprint statistic out of permutations of the locations in each chromosomal arm. Guttman et al. extended this method to scan a range of thresholds for defining aberrations, selecting multiple aberrations in each threshold (Guttman et al. 2007). Lipson et al. (2006) tried to identify optimal intervals over the aCGH data. Methods in similar spirit were developed for analysis of SNP data, which is informative for genotyping as well as copy number (Beroukhim et al. 2007;Weir et al. 2007).

Intuitively, an aberration is more likely to have biological significance if it happens in many samples, and if it is strong. A longer aberration is less likely to be attributable to measurement error. Thus, the three parameters used to score each marker are the number (or fraction) of carriers (patients), the length of the aberration and its amplitude. We refer to the product of these three factors as the volume $V$ of the marker, and use it as our statistic to assess the validity of each aberration. The method compares the real data to the randomized data obtained by permutations of the real data, under the null assumption that the genomic locations are independent. Once we obtain the distribution of $V$ in our randomized data, we can evaluate the statistical significance of the actual value of $V$, measured for each marker. We detect significantly aberrant genetic locations and associate them with a p-value. We demonstrate the method for three different public aCGH datasets from two different childhood neoplasms associated with the nervous system on three different BAC array platforms: Medulloblastoma – GSE8634; Neuroblastoma – GSE5784 (Tomioka et al. 2008) and GSE7230 (Mosse et al. 2007).

# Results

### Algorithm

Our method uses aCGH data to create a concise genomic description of each sample, including chromosomal status and appearance of significant local copy number aberrations. This concise description can be used to find an informative order or sub-classification of the samples.

The algorithm includes two steps – assigning chromosomal status and detecting significant local copy number aberrations. Amplifications and deletions are detected separately but similarly, using the same method.

### Input

The algorithm's input is the raw log2 aCGH data, and the markers' status. The raw log2 ratio data of chromosome 2p, taken from GSE7230, is presented in Figure 1A. Markers' status is the assignment per marker per sample – loss (-1), normal (0) or gain (1). The status was set by the R package GLAD (Gain and Loss Analysis of DNA) for identifying deleted or amplified genomic regions (Hupe et al. 2004) (see Supplementary note 1). Markers that were not correlated with their adjacent markers, but highly correlated with markers at another genomic location, were removed (see Methods, section 'Recognizing possible inaccurate genomic locations'). We constructed an amplifications matrix A, which has binary valued elements: $A_{ms} = 1$ if the aCGH marker m was assigned a gain value on sample s, and $A_{ms} = 0$ otherwise (the amplification matrix of chromosome 2p based on the GSE7230 data is shown in Figure 1B). A deletions matrix D is defined similarly: $D_{ms} = 1$ if the aCGH marker m has a loss assignment on sample s, and $D_{ms} = 0$ otherwise (deletion matrix is not shown). Markers' status is equal to A-D.

**Chromosome status**
We define an entire chromosome arm gain in a sample when more than 50% of the markers have a status of 'gain' in that sample. A sample in which an entire chromosome arm is lost is defined by more than 50% of the markers having a status 'loss'. For graphical representation of chromosomal status, the median log2 ratio of all markers on each chromosomal arm in each sample is used.

**'Volume' statistic**
Our goal is to find markers whose aberration happens significantly more frequently than expected by chance, taking into account the known tendency of cancer cells to gain and lose DNA sequences.

Three factors are relevant for assessing the significance of an aberration:

**Width W** - The number of carriers – the more tumors have an aberration, the more likely this aberration is to give selective advantage to the cell that carries it.

**Height H** - The amplitude of the aberration. Typically, a duplication event creates only one extra copy of the sequence. Thus, having multiple copies may indicate that having this amplification gives a selective advantage. This is more relevant for amplifications, as deletions can be only at two levels – hemizygous or homozygous deletion. In addition, the amplitude of the aberration measured in a certain tumor is affected by the fraction of subclones in the tumor tissue tested in which it is present. If the fraction is higher, the amplitude is higher. The amplitude is hard to compare among samples, as the range of values varies depending on the percent of diploid cells in the tumor sample.

**Length L** - The length of the aberration (number of neighboring markers included) is also important, but its contribution to the volume statistic defined below should be limited. The reason is that the aim of our analysis is to look for specific genes that "drive" the aberration, and long events, that affect the copy number of a large number of genes, are not informative. Therefore, entire chromosomal arm gains and losses are removed, and all the markers on this chromosome arm are given NaN value (arithmetic representation for Not-a-Number) for this sample. The removed chromosomal arms in each dataset appear in Supplementary Table 1, and their graphical representation in Supplementary Figure 1. If the statistic that characterizes the aberration increases linearly with the length, the presence of a few samples with very long aberrations can have a very strong effect on the results of the calculation. This can be avoided by setting on L an upper limit, denoted by K, and choosing K ~5-10 markers (in the actual implementation, we scan different values of K and combine the results). If the length of the aberration exceeds K markers, the value of the Length parameter is set to L=K. It should be noted that the number of markers does not necessarily reflect linearly the aberration length on the chromosome, as the distances between the markers are not uniform along the genome.

Our method takes into consideration all the three factors – width, height and length, in order to calculate the statistic termed 'volume' for each marker.

The detailed volume calculation is done as follows:

For each dataset there are two binary matrices – the amplification matrix A and the deletion matrix D, defined in the 'Input' section above. For samples in which an entire chromosome arm is gained (see 'Chromosome status' section above for definition), the corresponding entries of A are replaced by NaN, and for samples in which an entire chromosome arm is lost, the corresponding entries are replaced by NaNs in the deletion matrix D. Figure 1 displays the amplification volume calculation for chromosomal arm 2p in GSE7230 (Neuroblastoma). The height matrix H is actually the raw log2 ratio. $H_{ms}$ (Figure 1A) is the measured aCGH log2 ratio value of marker m in sample s. $A_{ms}$ (Figure 1B) is the amplification matrix, where each element (m, s) contains the digit 1 if the status of marker m on sample s is gain. In the length matrix L (Figure 1C), each element (m, s) containing the digit 1 in A is replaced by $L_{ms}$. $L_{ms}$ is the length of the sequence of ones on sample s, to which marker m belongs (length dimension). If $L_{ms} > K$, we set $L_{ms} = K$, to avoid overweighting long aberrations. In Figure 1 we used K=5. If $A_{ms} = 0$, $L_{ms} = 0$ as well. In the X matrix shown in Figure 1D, each element (m, s) containing the digit 1 (in A) is replaced by a real number $X_{ms}$, where $X_{ms} = H_{ms}*A_{ms}*L_{ms}$ ($A_{ms}$ is redundant here, as $L_{ms}=A_{ms}*L_{ms}$, and is included for clarity). Finally, all the numbers in row m are summed – representing the contribution of the width variable to our statistic $V_m$ (equation 1), representing the 'volume' of marker m (Figure 1E).

**Equation 1** $$V_m = \sum_s X_{ms}$$

This value is divided by the number of samples with non-NaN entries for this marker. This is done in order to enable detecting local aberrations even in genomic regions that are affected by large scale aberrations in most samples, but are nontheless containing some local aberrations. The volume statistic is calculated separately for each value of K, K = 1:10. Six markers are significantly amplified (significance threshold is marked by a red line in Figure 1E, see next section for details on setting the p-value per marker, and the FDR section in Methods for controlling the False Discovery Rate (FDR)). The raw aCGH data of these six markers are shown in Figure 1F, and their location is marked by a red asterisk in Figure 1A-E. The volume statistic is calculated separately for amplifications, using the amplifications matrix A, and for deletions, using the deletions matrix D.

**Associating p-values to the volume statistic of each marker**

Due to our lack of knowledge about the null distribution, in order to assign a p-value for the volume statistic of each marker m, $V_m$, a permutation of the original data is applied to approximate the distribution of the data under the null assumption of independence between the aCGH values and the genomic locations in different samples. In order to preserve the length distribution in each sample, we permute the X matrix (that already includes the length contribution to the volume statistic), and not the H or L matrices. This choice also saves recalculating the length of the aberrations. The entries of each column of the matrix X are permuted, and then the values in each row are summed. This randomization preserves the number of aberrant markers in each sample, their intensity, and the contributions of the lengths of the aberrations, while removing any location data. The randomization is repeated N (N=100) times (see 'Number of permutations' section below for discussion of robustness in N). Note that permuting across samples (rows) will have no effect on the computed volume. For each of the N randomized X matrices we calculate $V_i$ for every marker, obtaining for our n markers N*n values $V_i$. The distribution of these N*n numbers is used to calculate the p-value associated with every measured value of V, simply by counting the frequency of values in the null distribution that are higher or equal to the measured value.

The null distribution is estimated separately for each value of K from 1 to 10, for amplifications and deletions.

The FDR procedure (Benjamini and Hochberg 1995) was used to control the False Discovery Rates. See FDR section in Methods for details.

**Definition of an aberrant region**

After significantly aberrant markers are identified, adjacent markers, as well as markers separated by a single non-aberrant marker, are being combined into a single aberration. The aberration region is defined as the region between the non-aberrant markers that are bordering the aberration. Each aberration was annotated for being included in a normal copy number variation. In addition, genes residing within each aberration, and specifically cancer related genes, were listed (see Methods, section Aberrations' annotation).

**Parameters space**

**Maximal aberration length**

In order to avoid an overrepresentation of long aberrations, two measures were taken. First, for each chromosomal arm, in samples in which an arm status was 'gain' or 'loss', all marker values on this arm were replaced by NaNs. In addition, the maximal contribution of an aberration length to the volume was set to K. This K is an arbitrary value, representing preference to aberrations that are longer than one marker, but avoiding dominance of the signal by a few very long aberrations, which may induce ignoring short aberrations. Whenever an arbitrary value is assigned to a parameter, it's effect on the results has to be checked. As Supplementary Figure 2 shows, as the parameter K increases over the range 1-10, the number of significantly aberrant markers detected decreases monotonically, and the cumulative number of detected markers reaches a plateau. Therefore, we repeated all the analyses for K = 1:10.

**Number of permutations**

As we use the frequency of each volume in all permutations to assess the p-value, the more permutations there are, the more accurate is the result, as a frequency of zero will always be accounted for as significant. The number of permutations N thus may, in principle, affect the number of markers found significant. However, the actual distribution converges fast. Though the p-value of a given volume may vary a bit with increasing N, it reaches a plateau before N=100. For increasing N from 100 to 200, the change in the p-value for a given volume (corresponding to FDR of 0.1 or 0.01 for N=100) is smaller than $10^{-4}$. Thus, we chose to work with N = 100.

**Applications**

The method was applied to three datasets. Table 1 displays the number of aberrant markers and aberrations detected in each dataset. Significantly deleted markers appear in Supplementary Table 3, and deletions in Supplementary Table 5. Significantly amplified markers appear in Supplementary Table 4, and amplifications in Supplementary Table 6.

**Medulloblastoma**
When applied to the Medulloblastoma dataset analyzed here (GSE8634) our method finds all the known chromosomal aberrations of this cancer, and several possibly new ones as well. Figure 2 displays the chromosome status map of the Medulloblastoma dataset, and the significant aberrations. As described in GSE2139 (Mendrzyk et al. 2005), where a subset of the samples were analyzed, isochromosome 17 (i(17q) - loss of 17p, replaced by an exact copy of 17q) is the most frequent aberration. We identified five different subgroups by manually ordering the samples, marked on the bar below Figure 2 – Subgroup 1 has many chromosomal aberrations, but not isochromosome 17. Subgroups 2 and 3 carry isochromosome 17, which is the most frequent aberration in Medulloblastoma (Mendrzyk et al. 2005). On the basis of our analysis, we propose that the tumors displaying this aberration can be further separated into a group with many chromosomal events (marked 2) and a group with no other common chromosomal events (marked 3). Group 2 is analogous to Neuroblastoma type 1, one of the three clinicogenetic subgroups described in Neuroblastoma (Brodeur 2003;Vandesompele et al. 2005), in the sense that there are many events of loss and gain of chromosomal arms that are common to the samples in this group. Several events of gain of chromosome 7 in group 2 are accompanied by loss of 8, resulting in chromosomes 7 and 8 being negatively correlated. A subgroup of tumors with loss of chromosome 6 (marked 4, genomically characterized similarly to (Kool et al. 2008) cluster A, associated with WNT and TGFβ signalling) do not have isochromosome 17, as described also in other Medulloblastoma datasets (Clifford et al. 2006;Thompson et al. 2006). The last group (marked 5) has few or no chromosomal events. Three tumors of that group have gain of chromosome 7, and three samples have loss of chromosome 22, but those numbers are too small to consider them as separate subtypes. It would be of interest to compare this chromosomal status-based stratification of Medulloblastoma to previously defined subgroups,, such as SHH associated and WNT associated (Thompson et al. 2006). However, this cannot be done since the present dataset is not annotated clinically. This classification only partially corresponds to the partition of (Kool et al. 2008), because their partition was based on gene expression.

Our method identified 10 amplified regions (Supplementary Table 6A) comprised of 13 amplified markers (Supplementary Table 4A), and 99 deleted regions (Supplementary Table 5A) comprised of 137 deleted markers (Supplementary Table 3A). Figure 2B displays selected aberrations. MYCN and CDK6 amplifications were identified. MYCN region amplification appears only in groups 1-3. Amplification of the CDK6 region appears mostly in groups 1 and 2. NPM1 (Nucleophosmin, B23) was deleted in few samples. NPM1 has been recognized as a partner gene for various chromosomal translocations in hematological malignancies. NPM1 was associated with centrosome duplication and the regulation of p53, and might have a role as a tumor suppressor (cf Naoe et al. 2006).

This dataset (GSE8634) has not yet been published, but dataset GSE2139 that includes a subset of the samples (Mendrzyk et al. 2005) was analyzed for local aberrations. This publication included a list of amplifications and deletions. We searched for markers that were included in amplifications or deletions identified there and by our method. Three of the amplifications reported there included markers that were identified as significantly amplified by our method – MYCN, CDK6 and marker RP11-382A18. Marker RP11-382A18 is annotated near MYC region on chromosome 8q by the platform of GSE2139, used by (Mendrzyk et al. 2005). MYC amplification and MYCN amplification are mutually exclusive. Nine of the amplifications reported there were not identified by our method. Four of their deletions included markers that were identified as significantly deleted by our method, annotated there to carry CHRD, UTF1, PRDM2 and HDAC4. Eight of the amplifications reported by (Mendrzyk et al. 2005) were not identified by our method.

**Neuroblastoma**

Figure 3 displays the chromosome status map of both Neuroblastoma datasets, as well as the aberrations common to the two Neuroblastoma datasets tested. Samples are manually ordered according to the three distinct clinicogenetic subgroups described in Neuroblastoma (Brodeur 2003;Vandesompele et al. 2005). The first group (marked 1 on the bar below Figure 3 subplots) exhibits predominantly full chromosomal aberrations (typical gains of chromosomes 6, 7, and 17, and losses of chromosomes 3, 4, 11, and 14). Both other two groups (marked 2A and 2B) are characterized by structural chromosome aberrations, such as partial 17q gain. Group 2A has MYCN amplification and 1p deletion. Group 2B is characterized by 11q deletion, and to a lesser extent, 3p deletion. This classification explains most of the chromosomal arms associations found.

In GSE5784 there are 15 amplifications (Supplementary Table 6B, 28 markers amplified, Supplementary Table 4B) and 115 deletions (Supplementary Table 3B, 245 markers deleted, Supplementary Table 5B). In GSE7230 there are 18 amplifications (Supplementary Table 6C, 30 markers amplified Supplementary Table 4C) and 49 deletions (Supplementary Table 5C, 87 markers deleted, Supplementary Table 3C). Three amplifications and 14 deletions are common to both Neuroblastoma datasets (GSE5784, GSE7230) (Table 2, Figure 3 C and D). The first amplified region, which was separated into two regions in GSE7230, is on chromosome 2, and corresponds to the MYCN region. MYCN amplifications were identified mostly in group 2. The other amplification is of the defensins cluster on chromosome 8. In addition to being amplified in several samples, this region is deleted in other samples, in accordance with this region being a known frequent normal copy number variation (Hollox et al. 2003). Eight of the common deletions correspond to the 1pter deletion, and this deletion was fractioned into eight deletions in GSE7230. Another common deletion is in the region of BRCA1, a known tumor suppressor gene.

In GSE5784, several known tumor suppressor genes were deleted - APC, CDKN2A, RB1 and TGFBR1. Also, two regions with known oncogenes were amplified in this dataset - a region on chromosome 11, that includes CCND1, FGF19, FGF3, FGF4 was amplified, as well as a region on chromosome 12 with ETV6. For GSE5784, no aberration list was given in the original publication (Tomioka et al. 2008) for comparison.

In GSE7230, the ALK region on chromosome 2 was amplified. ALK was previously identified as having a role in Neuroblastoma (Osajima-Hakomori et al. 2005). The fumarate hydratase (FH) region was deleted in GSE7230. FH was shown to be a tumor suppressor gene in several cancers (Tomlinson et al. 2002). For GSE7230 (Mosse et al. 2007) aberrations are reported at the cytoband level. Only two of the 24 amplified regions that were reported overlap with amplifications identified by our method – MYCN and a region on chromosome 16. Eleven of the 22 deleted regions reported in (Mosse et al. 2007) overlap with deletions identified by our method, including the 1pter deletion and MLH1 region.

## Discussion

We have introduced a simple intuitive method to recognize significant local amplifications and deletions in aCGH data. The input is the raw data, and its categorization into gain, normal and loss values for each marker in each sample (defined in our implementation by GLAD (Hupe et al. 2004)). Then, for each marker, its level of change, frequency of change and length of change are combined to create a volume statistic. The significance of this statistic is assessed using a random distribution based on a permutation of all the data. After aberrant markers are detected, they are combined into continuous aberrations that are annotated for normal copy number variations and then associated with cancer related genes.

**Parameters' dependence**

Our guiding principle was to keep the method simple. We wanted to incorporate as few assumptions and as few arbitrary parameters as possible into the method. Implementation of the method necessitates setting three parameters: number of randomizations N, maximal aberration length contribution for statistic calculation, K, and FDR level. The number of permutations N affects the computation time. As the distribution of the volume statistic under permutations converges fast, increasing N above 100 will not change the results.

The value of K, the maximal aberration length contribution for the statistic used, does affect the identity of the aberrations detected as significant. Thus, we scanned for K = 1:10, and combined the results. We showed that increasing K above 10 had very little effect on the aberrations detected.

The chosen FDR level naturally affects the results, but setting the level of acceptable false discovery rate, the multiple comparisons equivalent of the confidence, is always left to the researcher to decide. However, the minimal volume required for an aberration to be detected as significant at each level of FDR can be estimated per each value of K, and the FDR level can be adjusted accordingly.

**Statistic calculation**

There is no reason to assume that the number of carriers, length and amplitude of an aberration are equally important to set its significance, as they are used here to calculate the 'volume' statistic. But they are all biologically relevant parameters, and lacking an educated weighting system for these parameters, this is the simplest way. The relative weight of each parameter can be easily changed within this framework. Actually, we vary the relative weight of the length parameter when varying K. We also tested the case where the Height parameter is ignored, but this causes the loss of detection of relatively rare strong amplifications (eg CDK6 amplicon in Medulloblastoma).

**Status assignment**

The accuracy of status assignment (gain/loss/normal) may affect on the results. If thresholds are too restrictive, aberrant markers may not be recognized as such. If thresholds are too permissive, many markers will be considered as aberrant. This may hamper the ability of the method to identify weak or rare aberrations. There are several methods for status assignment available today (Fridlyand et al. 2004;Hupe et al. 2004;Myers et al. 2004;Olshen et al. 2004;Eilers and de Menezes 2005;Hsu et al. 2005;Lingjaerde et al. 2005;Picard et al. 2005;Wang et al. 2005), and the user may select the method most appropriate for his data.

**Normal copy number variations**

The normal copy number variation is a complicated issue in detecting significant disease related aberrations. Discarding all aberrations that contain any known variation will remove most of the aberrations, including clinically recognized ones. In addition, the normal copy number variation database contains variations that were identified on patients with various medical conditions that may affect copy number. Thus, only variations identified on normal population on a similar platform (Redon et al. 2006) were used for annotation. In addition, every marker that was both significantly deleted and significantly amplified was recorded as suspected for normal copy number variation. Indeed, many significantly aberrant locations are annotated as frequent normal copy number variation. In cases when there are enough normal and tumor samples of the same population, it may be interesting to see how significantly the frequencies of high or low copy numbers of certain normal copy number variations differs between the normal and tumor populations, which may serve as an indication for a possible predisposition of carriers of those variants to cancer.

**Problematic marker annotations**

Another problem of most aCGH platforms is problematic marker annotations. In clustering the markers on the basis of their aberration profile for each dataset, up to 5% clustered with markers annotated to other chromosomes (data not shown). This is an under-estimation of the number of wrongly annotated markers, as not all chromosomes create a stable cluster of the associated markers. This problem can be addressed in several ways. The simplest one is to discard all single marker aberrations. This however may result in losing valuable information. Thus, we removed markers that had low correlation with the chromosome to which they were assigned and high correlation with another chromosome. Still, many of the significantly aberrant markers are not correlated to their adjacent markers, and are still suspected to be located elsewhere in the genome.

**Treating long aberrations**

Unlike previous works, we do not perform binning into fixed–width locations that may incorporate artefacts (Diskin et al. 2006). The volume statistic we use is similar to the frequency statistic used in (Diskin et al. 2006) for k= 1, i.e. when the aberration's length is not taken into account to calculate its significance. Another difference is that we compare each marker to all the genome, and not to a certain chromosomal arm, thus applying an equal 'significance' threshold to all aberrations. To enable this, all long events must be removed. In most cases, removing all chromosomal arms on which more than half of the markers are aberrant, is enough. However, in certain cases (11p in both cancers, 1p in Neuroblastoma) we noted long events of less than half an arm length that were not removed. When these events are on the same genomic location, they may cause identification of many markers in this region as aberrant, always in the same samples. This may be correct, but is not the goal of this analysis, aimed at finding local aberrations. Thus, in such cases, long chromosomal events can be noted and removed prior to the analysis, or after the analysis. Removing these aberrations, that may be interesting in themselves, may allow the detection of more local aberrations.

**Biological findings**

When comparing the aberrations identified by our method to the aberrations identified by other methods, we see all the oncogenes that are known to be amplified in the corresponding cancers, but our method misses some aberrations identified in previous publications and finds new one. This is a natural consequence of the parameters we defined and the removal of whole arm events. One of the main differences we have, using our method, is that identification of a region that is aberrant in one sample only as significant is rare. Also, a region that is amplified on an amplified chromosome background, or a region that is deleted on a deleted chromosome background with not many separate appearances on a normal copy number background cannot be identified, as chromosome level events are removed. This is in agreement with our goal of detecting local events. However, this can be overcome by running the method for each chromosome or chromosome arm separately, which would allow inclusion of all samples in the calaculation, and identification of local amplifications on the background of chromosomal amplifications, and of local homozygous deletions on the background of chromosome loss. However, that approach also has several drawbacks. First, in aCGH with several thousands markers, the number on markers on some of the smaller chromosomes is too small to allow for generation of a reliable null distribution. Second, it is difficult to find short non frequent aberrations on chromosomes that have long events, and third, the threshold an aberration has to pass to be considered as significant will be different for different chromosomes. In cases where the goal is to find an exhaustive list of aberrant locations, one may consider applying the analysis for the entire genome and for each chromosomal arm separately. We applied our method on three public datasets of childhood neoplasms associated with the nervous system - one of Medulloblastoma (GSE8634) and two of Neuroblastoma (GSE5784, GSE7230). In Medulloblastoma, we find five distinct sub groups. Two sub groups with isochromosome 17, one with many other chromosomal events (2), and one with few chromosomal events (3). There is also a group with many chromosomal aberrations but without isochromosome 17 (1), a group with loss of chromosome 6 (4), and a group with few aberrations (5). MYCN amplification appears only in the first three groups, and CDK6 amplification appears mostly in the first two types. MYC amplification appears only when there is no MYCN amplifications, and only in the first two types, strengthening our new suggested partition of the isochromosome 17 type into two subtypes, the first of which is equivalent to Neuroblastoma type 1.

In Neuroblastoma, we identified the three known subgroups, and the MYCN amplification known to be associated with one of the types.

Comparing two types of childhood neoplasms associated with the nervous system, it is interesting to note the role of chromosome 17, and its interrelations with MYCN amplifications. Chromosome 17 amplification has two forms – isochromosome in Medulloblastoma, and gain of the q-arm or the whole chromosome in Neuroblastoma. MYCN amplification appears mostly with isochromosome 17 in Medulloblastoma, but only with 17q amplification in Neuroblastoma – rarely with whole chromosome gain. It was recently shown that MYCN-directed centrosome amplification, leading to increased tumorigenesis, requires MDM2-mediated suppression of p53 activity in Neuroblastoma cells (Slack et al. 2007). Since p53 is located on chromosome 17p, it can be suggested that suppression of p53 is difficult when there are more than two copies of 17p, and thus there is no selective advantage in MYCN amplification in tumors carrying more than two copies of the full chromosome 17 (Neuroblastoma type 1). Similarly, MYCN amplification is more advantageous if there is deletion of 17p, carrying p53.

# Conclusions

Our method allows for a fast and biologically motivated detection of aberrant chromosomal regions, and associates them with chromosomal arm level events to characterize subtypes of cancer. We believe that our method is conceptually simpler to understand than prebiously published methods. We have demonstrated the ability of the method to detect all the clinically relevant proven aberrations and new DNA amplifications and deletions in two types of childhood neoplasms associated with the nervous system. In addition to the known chromosomal aberrations and known subgroups, our method identified a new subgroup in Medulloblastoma.

# Methods

**Datasets**

All aCGH datasets used for analysis were downloaded from GEO (see Table 1). Log2 ratios were used as appeared in GEO. Markers were ordered by their genomic location according to the annotation of the corresponding platforms. Loss, normal or gain status was assigned per each marker in each sample by GLAD (Hupe et al. 2004), using the parameters as are used in the GLAD manual (Supplementary note 1).

**Aberrations' annotation**

Normal copy number variations were downloaded from http://projects.tcag.ca/variation/ for the human genome versions hg17 and hg18. There is no data for hg16. Aberration annotation includes variations identified by aCGH on 270 normal individuals (Redon et al. 2006).

The list of genes in each aberration was created based on the genomic location from UCSC matched version knownGene table, gene symbols by kgXref table. The genes list in each aberration was scanned to search for cancer related genes ((Futreal et al.

2004), October 30, 2007 version). The lists of deletions and amplifications and their associated genes and cancer related genes appear in Supplementary Tables 5 and 6, respectively.

**Recognizing possible inaccurate genomic locations**

Our working hypothesis is that at least 90% of the markers are annotated to their correct chromosomal locations. In order to identify markers that we suspect to be mistakenly annotated, we use the correlation of the marker's signal intensity with neighboring markers. If the signal of a marker is correlated to that of its neighbors, it is not likely to be inaccurately annotated. Thus, we calculated for each marker m its Pearson correlation coefficients, c(m-1,m) and c(m,m+1), to its two neighboring markers. We define a threshold T for each dataset such that 20% of the correlations between adjacent markers are lower than T (assuming that less than 10% of the markers are inaccurately located, results in at most 20% of the neighbors being incorrectly identified as such). For most markers m, both c(m-1,m) and c(m,m+1) > T. A low correlation (less than T) to one of the two neighbors may be due to chromosome arm start or end, to an aberration or variation border, or to mistaken location annotation of the neighbor. If, however, the correlation of a marker to both its neighbors is below threshold, it is likely to be on an isolated aberration (copy number change) or - inaccurately located. We flagged these markers as suspected as being assigned to wrong locations. For each suspected marker we applied the procedure described in Supplementary Note 2, to check whether it can be confidently assigned to another genomic location, based on a very high correlation to the aCGH values of several markers in the other genomic location. If so, it was removed (see Supplementary Table 2 for lists of removed markers, together with their putative correct chromosomal arm), otherwise it was left in the analysis.

Potentially inaccurate location was identified for 17 to 144 markers per dataset, which constitute 0.7 – 3.5% of the markers (see Table 1).

We noticed for GSE8634 that many aberrations were highly correlated, and correlated to gender. Some of the samples were probably hybridized to opposite sex control samples. The 28 markers whose two sided t-test p-value between the genders passed FDR of 1% were thus removed, and the analysis was repeated. We assume those markers are actually located on the gender chromosomes, but as no data is included for markers on the gender chromosomes, we used the gender annotation.

**FDR**

Whenever many comparisons are done in parallel, p-values can be adjusted to control an overall error criterion. Here, we controlled the expected rate of false identifications of aberrations through the FDR criterion, as defined by Benjamini and Hochberg (1995). This procedure was applied on permutation p-values in Reiner et al (2003) and was shown there to control the FDR, based on simulated data.

The FDR controlling procedure was applied on the p-values of all markers from all chromosomes. The designated rate of false discoveries q will naturally affect the number of markers identified as significant, and should be set according to the dataset and the resources allocated to check significant aberrations. When q is set higher, the list of markers that are found significant is longer, but the expected rate of false positives also increases. In the analysis presented in this article, FDR was controlled at the 5% level.

The output of the algorithm is a table with ten columns, one for each value of K (K=1:10), and a row for each marker. FDR was controlled for the whole table, and only rows in which at least one column was found significant were defined as aberrations.

## Authors' contributions

TS and ED conceived the study. TS carried out the analysis. AR consulted the model and statistical analysis. WLL and MEH formulated the biological motivation. TS and ED drafted the manuscript. All authors read and approved the final manuscript.

## Acknowledgements

E.D. and T.S. were supported in part by the Ridgefield Foundation and the European Commission (EC FP6), and by a Program Project Grant from the National Cancer Institute (P01-CA65930).

# Figures

**Figure 1 - Calculation of the "volume" statistic for chromosomal arm 2p amplifications in GSE7230 (Neuroblastoma)**

(A) The height matrix H (raw data) of 2p, where each element (m, s) on 2p is the log2 ratio of aCGH marker m in sample s. Each row corresponds to a marker, and each column corresponds to a sample. For presentation only, values are truncated to [-1,1]. (B) The amplifications matrix A, where each element (m, s) on chromosome 2p that is amplified in sample s is marked by 1, otherwise 0. (C) The length matrix L of 2p, where each element (m, s) on chromosome 2p for which $A_{ms}=1$ is replaced by the length of the sequence of 1s to which it belongs on sample s. Maximal represented length is K=5. Non amplified markers are white. (D) X, the matrix created by multiplying elements of H, A and L. Non amplified markers are white. (E) Averaging the rows of X gives the volume statistic. The red line is the value of the volume statistic above which it is significantly amplified (corresponding to FDR of 0.05). (F) The markers of the only region on chromosome 2p that passes this threshold – the MYCN region, marked in A-E by red asterisks. For presentation only, values are truncated to [-1,1].

**Figure 2 - Chromosomal status and aberrations in Medulloblastoma**
(A) Chromosomal status of datasets GSE8634. Each row corresponds to a chromosomal arm. Due to space limitation, only every second arm is labelled. Since some chromosomes are telocentric (with short p arm), there is a change from p to q. Values are color coded according to the mean log2 ratio of the markers on each chromosomal arm. (B) Discussed aberrations in Medulloblastoma dataset GSE8634. Each column corresponds to a sample. Samples are manually ordered according to known and new clinicogenetic subgroups, as the bar below shows. Each row corresponds to an aberration discussed in the text, and the label indicates the gene associated with it. Values are color coded according to the mean log2 ratio of the markers on each aberration. In all subfigures, for presentation only, values are truncated to the range [-1, 1], rising from blue to red.

**Figure 3 - Chromosomal status and aberrations common to both Neuroblastoma datasets**
Chromosomal status of datasets GSE5784 (A) and GSE7230 (B), and the aberrations common to both of Neuroblastoma datasets, shown for the patients of GSE5784 (C) and GSE7230 (D). Each column corresponds to a sample. Samples are manually ordered according to known and new clinicogenetic subgroups, as the bar below shows. In A and B, each row corresponds to a chromosomal arm. Due to space limitation, only every second arm is labelled. Since some chromosomes are telocentric (with short p arm), there is a change from p to q. Values are color coded according to the mean log2 ratio of the markers on each chromosomal arm. In C and D, each row corresponds to a common aberration, and the label indicates the chromosome on which the aberration resides. Values are color coded according to the mean log2 ratio of the markers on each aberration. In all subfigures, for presentation only, values are truncated to the range [-1, 1], rising from blue to red.

## Tables

**Table 1 - Array CGH datasets analyzed**

| Dataset | Condition | Samples # | Markers # | Markers amplified | Amplifications | Markers deleted | Deletions | CNV | Removed # |
|---|---|---|---|---|---|---|---|---|---|
| GSE8634 | Medulloblastoma | 80 | 6295 | 13 | 10 | 137 | 99 | 4 | 126 |
| GSE5784 | Neuroblastoma | 236 | 2457 | 28 | 15 | 245 | 115 | 4 | 17 |
| GSE7230 | Neuroblastoma | 82 | 4073 | 30 | 18 | 87 | 49 | 0 | 144 |

The datasets are recognized by their Gene Expression Omnibus (GEO - [www.ncbi.nlm.nih.gov/geo](www.ncbi.nlm.nih.gov/geo)) series ID. CNV is the number of markers found to be significantly deleted and significanty amplified. Removed # is the number of markers removed prior to analysis because of a probable wrong annotation.

**Table 2 - Aberrations common to both Neuroblastoma datasets**

|  | GSE5784 | GSE5784 | GSE7230 | GSE7230 |  |
|---|---|---|---|---|---|
| chromosome | start marker | end marker | start marker | end marker | interesting genes |
| amplifications | | | | | |
| 2 | H10_K5 | H10_M34 | CTD-2603D17 | CTD-2603D17 | |
| 2 | H10_K5 | H10_M34 | RP11-775D5 | RP11-149C19 | MYCN;NAG; |
| 8 | H9_L19 | H9_I19 | RP11-499J9 | RP11-499J9 | defensins |
| | | | | | |
| deletions | | | | | |
| 1 | H11_N30 | H11_C10 | RP11-82D16 | RP11-780N18 | TP73; |
| 1 | H11_N30 | H11_C10 | RP11-327P18 | RP11-327P18 | |
| 1 | H11_N30 | H11_C10 | RP11-150L14 | RP11-707I5 | |
| 1 | H11_N30 | H11_C10 | RP11-728G12 | RP11-728G12 | |
| 1 | H11_N30 | H11_C10 | RP11-155L18 | RP11-155L18 | |
| 1 | H11_N30 | H11_C10 | RP11-598N19 | RP11-598N19 | |
| 1 | H11_N30 | H11_C10 | RP11-335G20 | RP11-335G20 | |
| 1 | H11_N30 | H11_C10 | RP11-219O7 | RP11-219O7 | |
| 4 | H9_C33 | H9_A3 | RP11-358C18 | RP11-358C18 | |
| 6 | H9_J12 | H9_J12 | CTD-2356O12 | CTD-2356O12 | |
| 7 | H11_M23 | H11_M22 | RP11-32H11 | RP11-32H11 | |
| 11 | H11_A33 | H11_O18 | RP11-367J12 | RP11-367J12 | |
| 17 | H11_J19 | H11_J19 | CTD-2321N2 | CTD-2321N2 | |
| 17 | H10_A32 | H10_A32 | CTD-2321N2 | CTD-2321N2 | |

# Additional files

**Additional file 1 – Supplementary information**

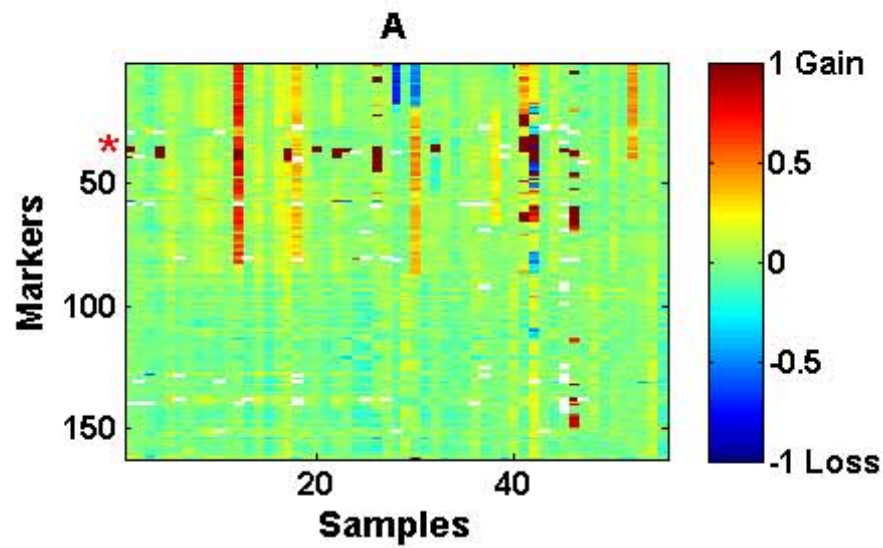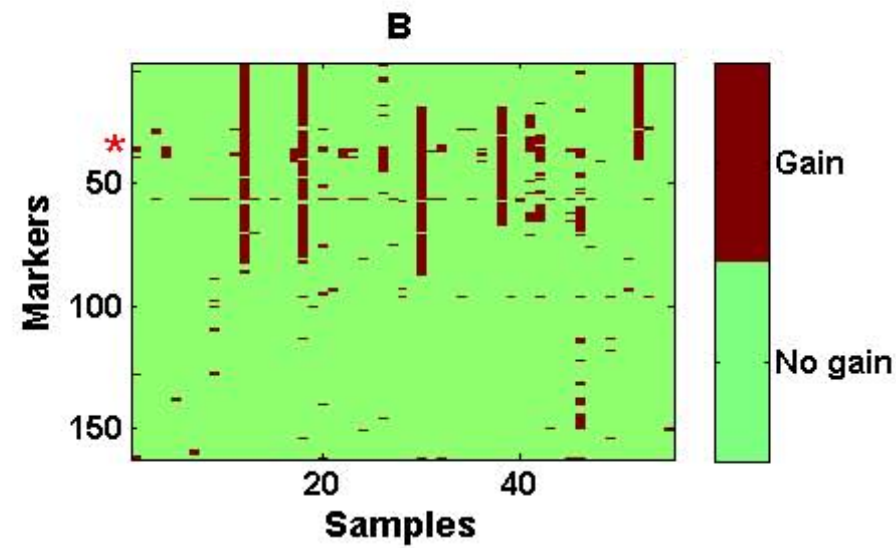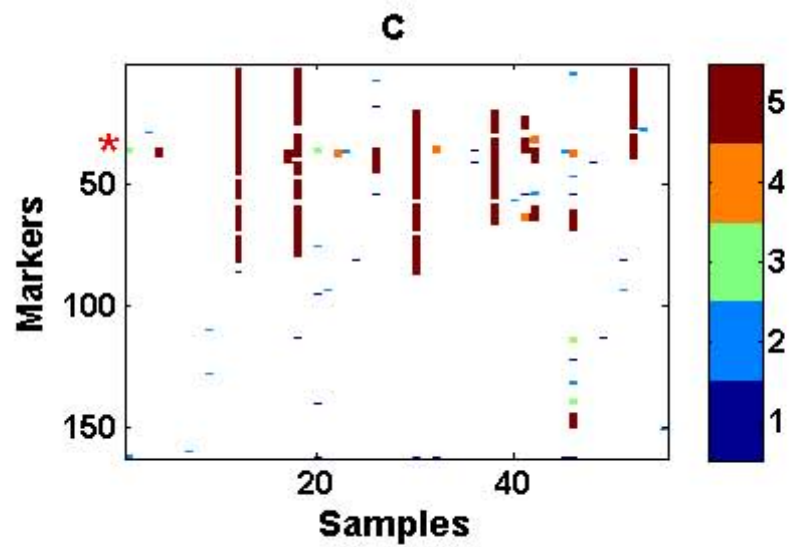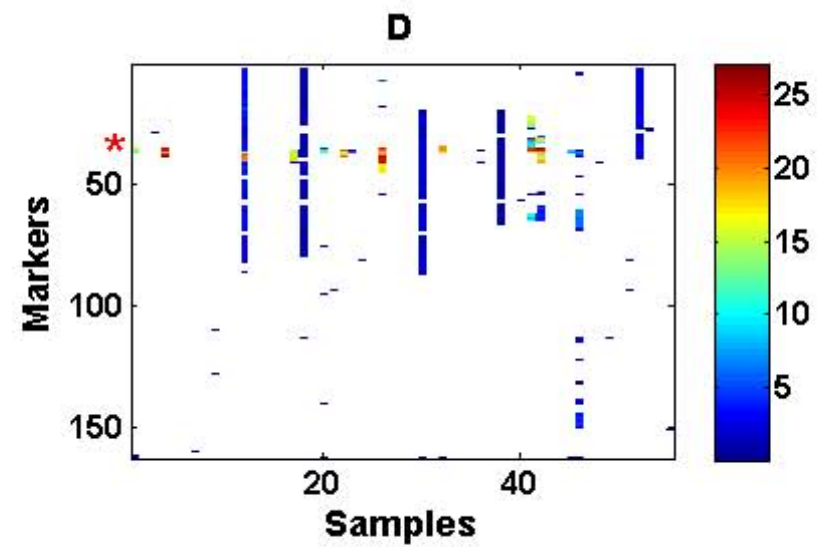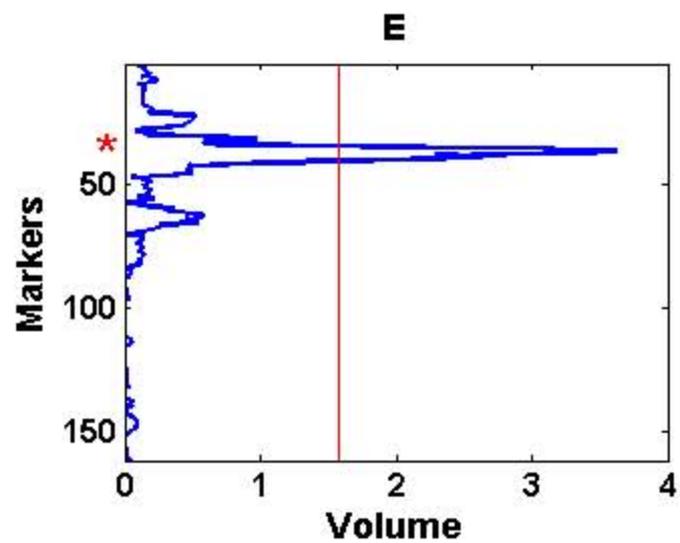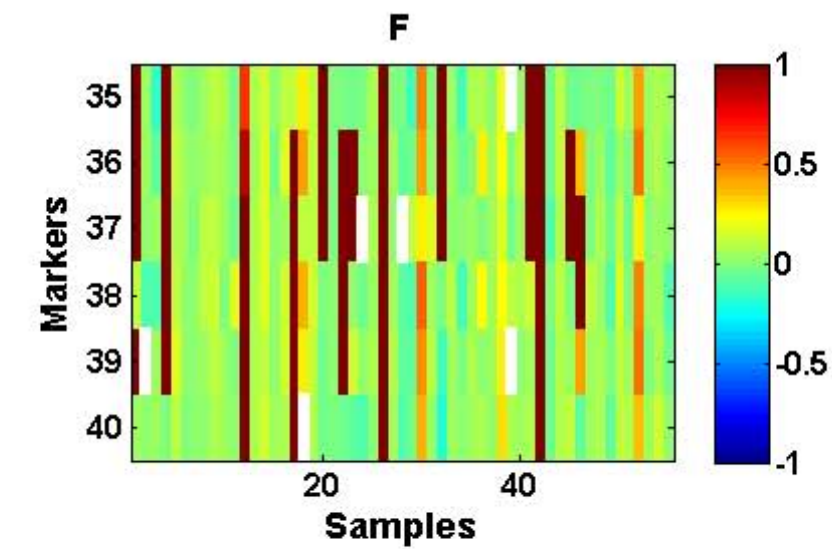

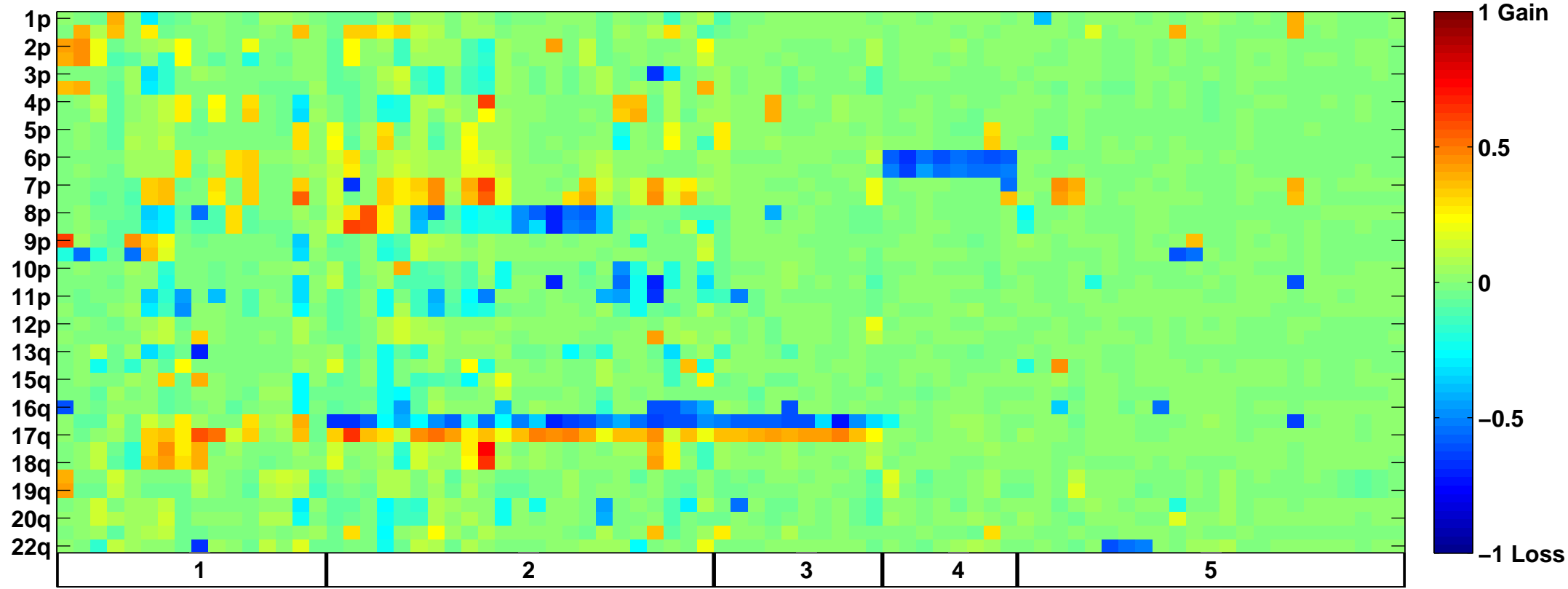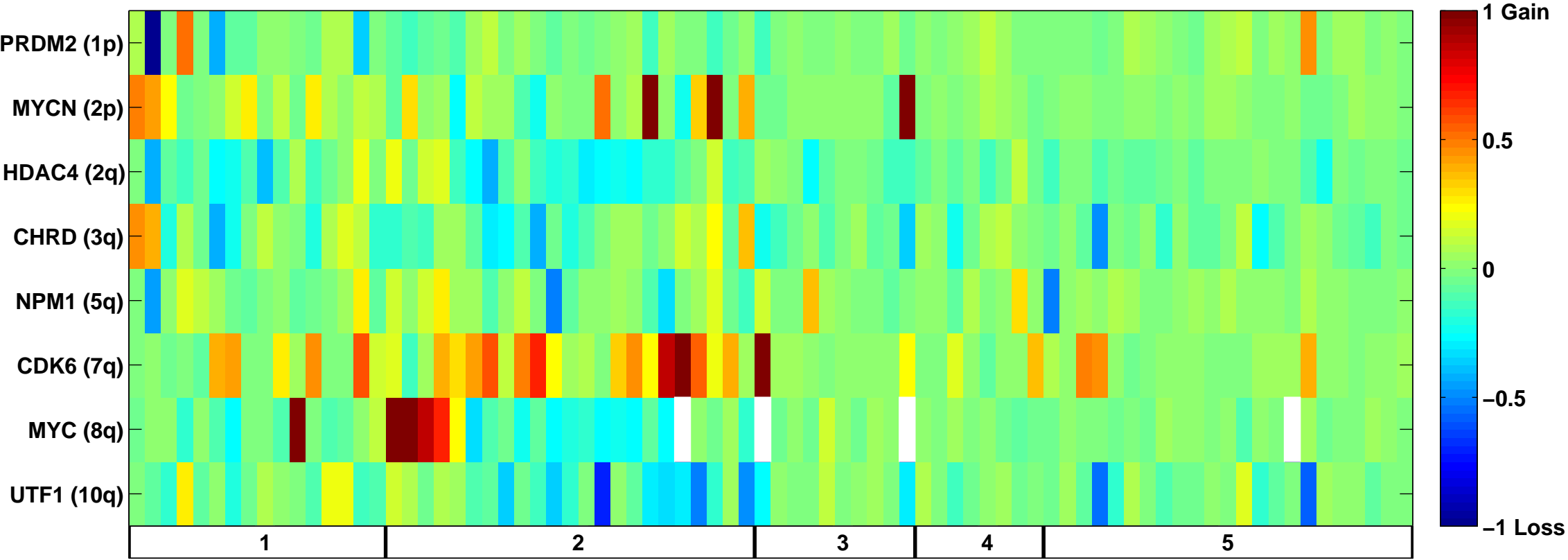

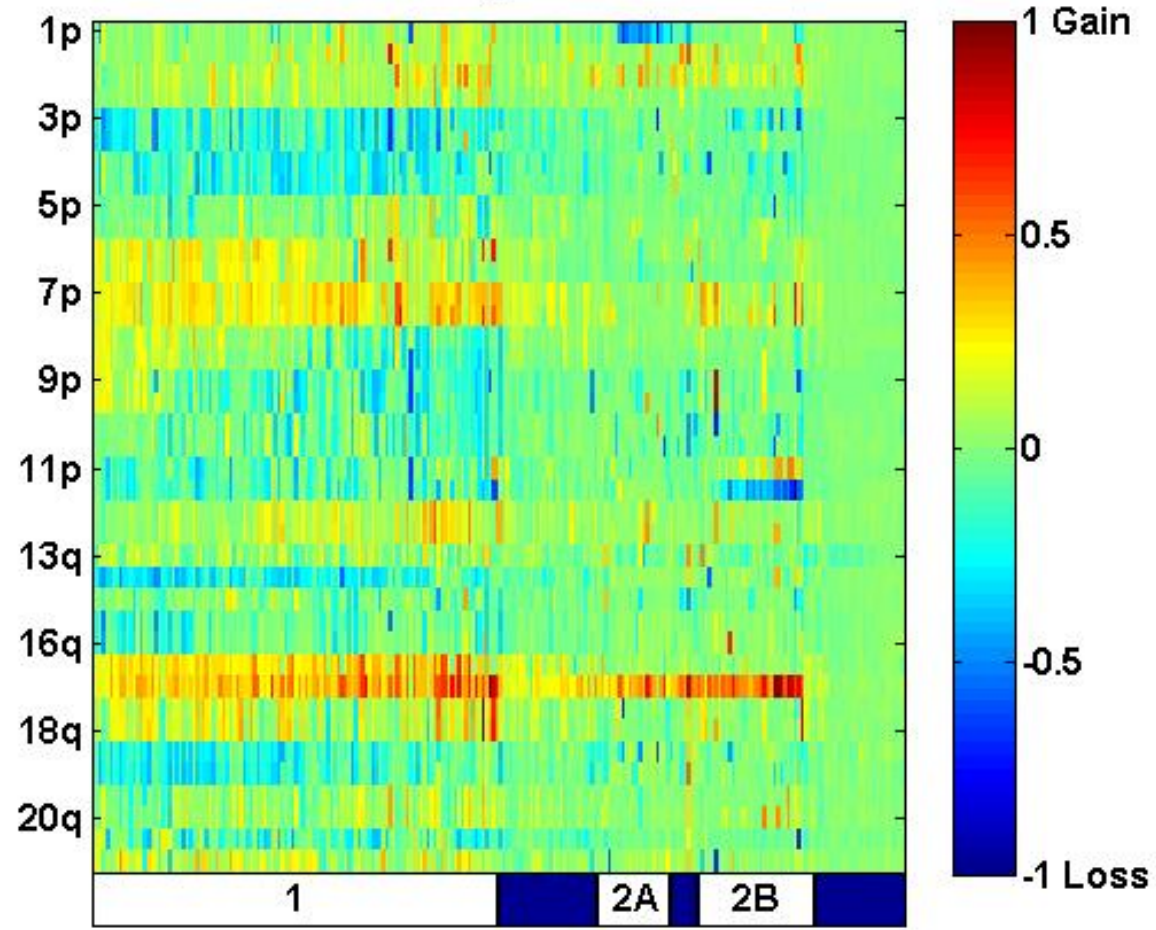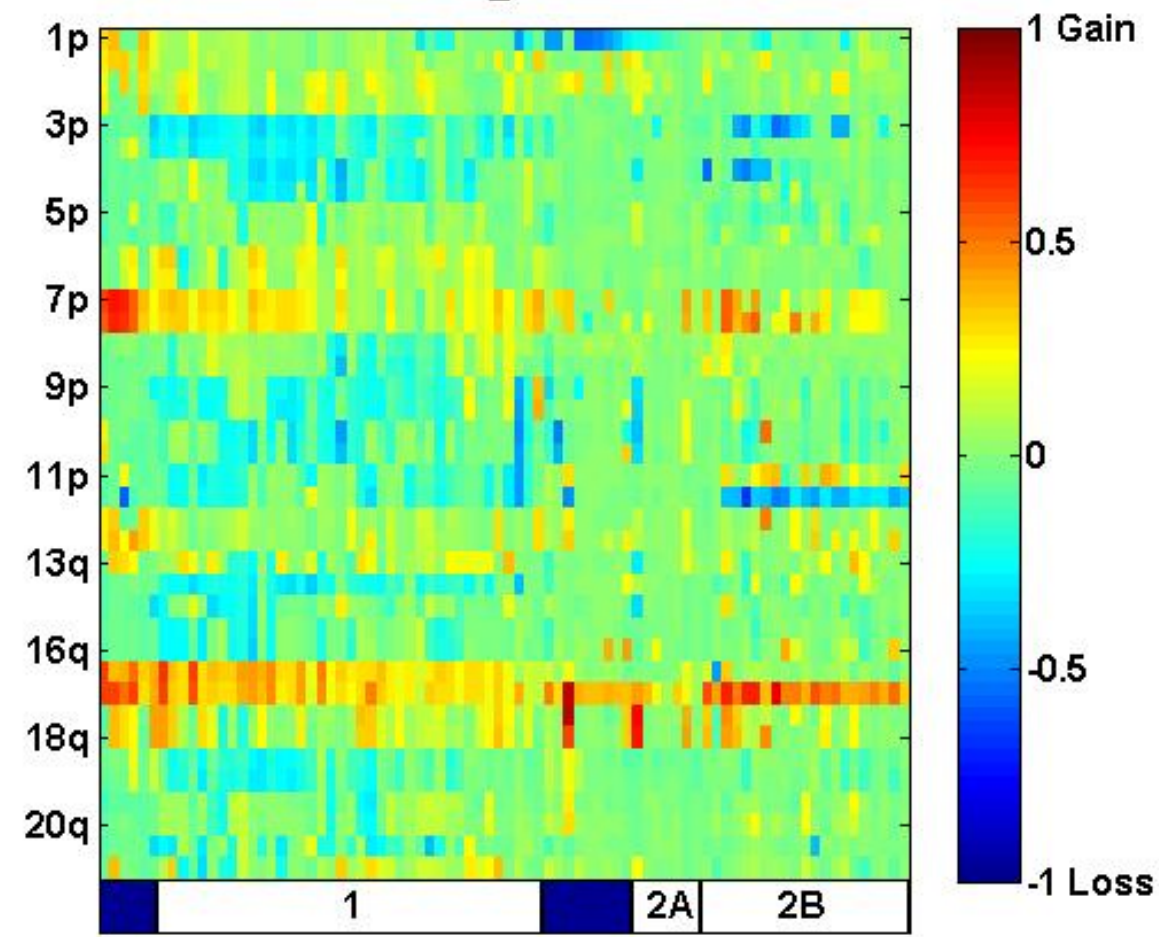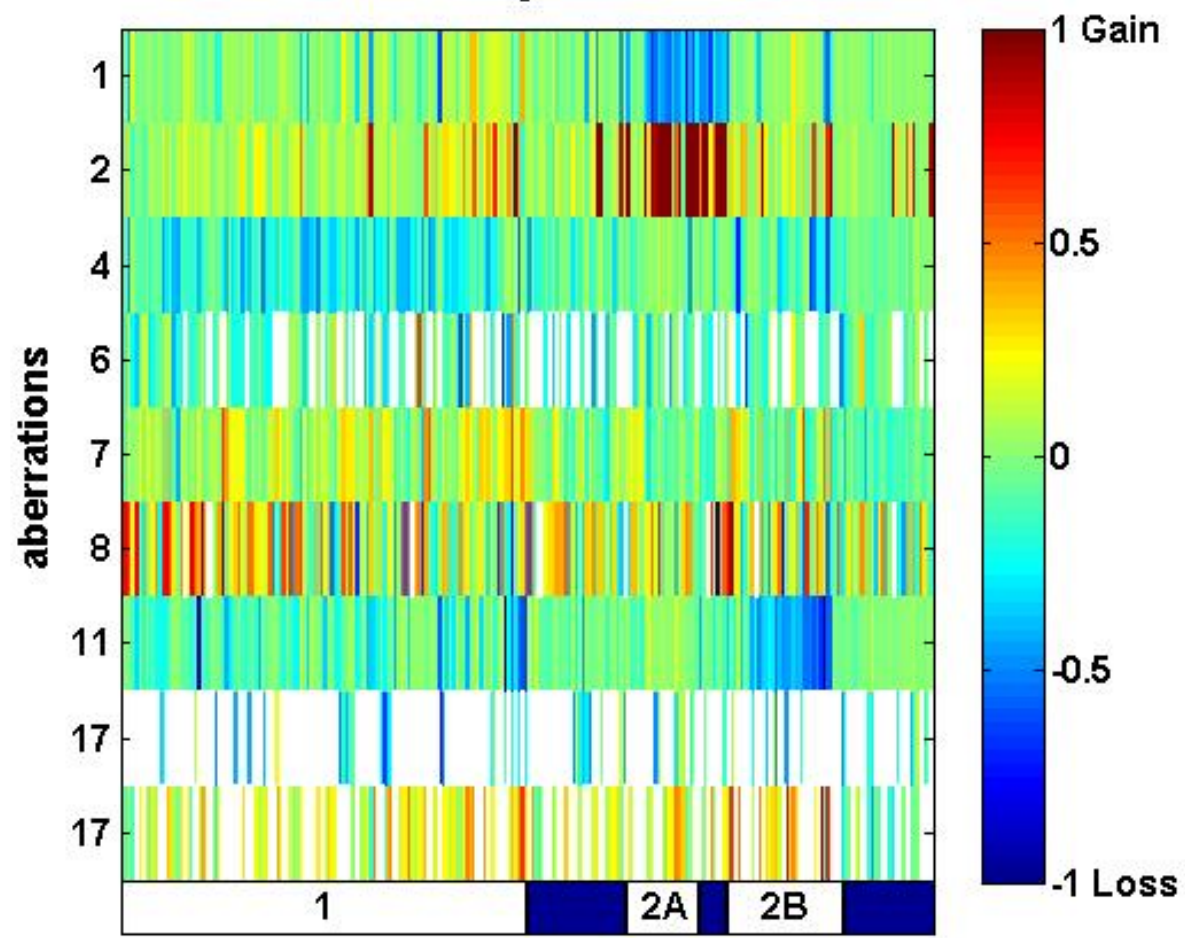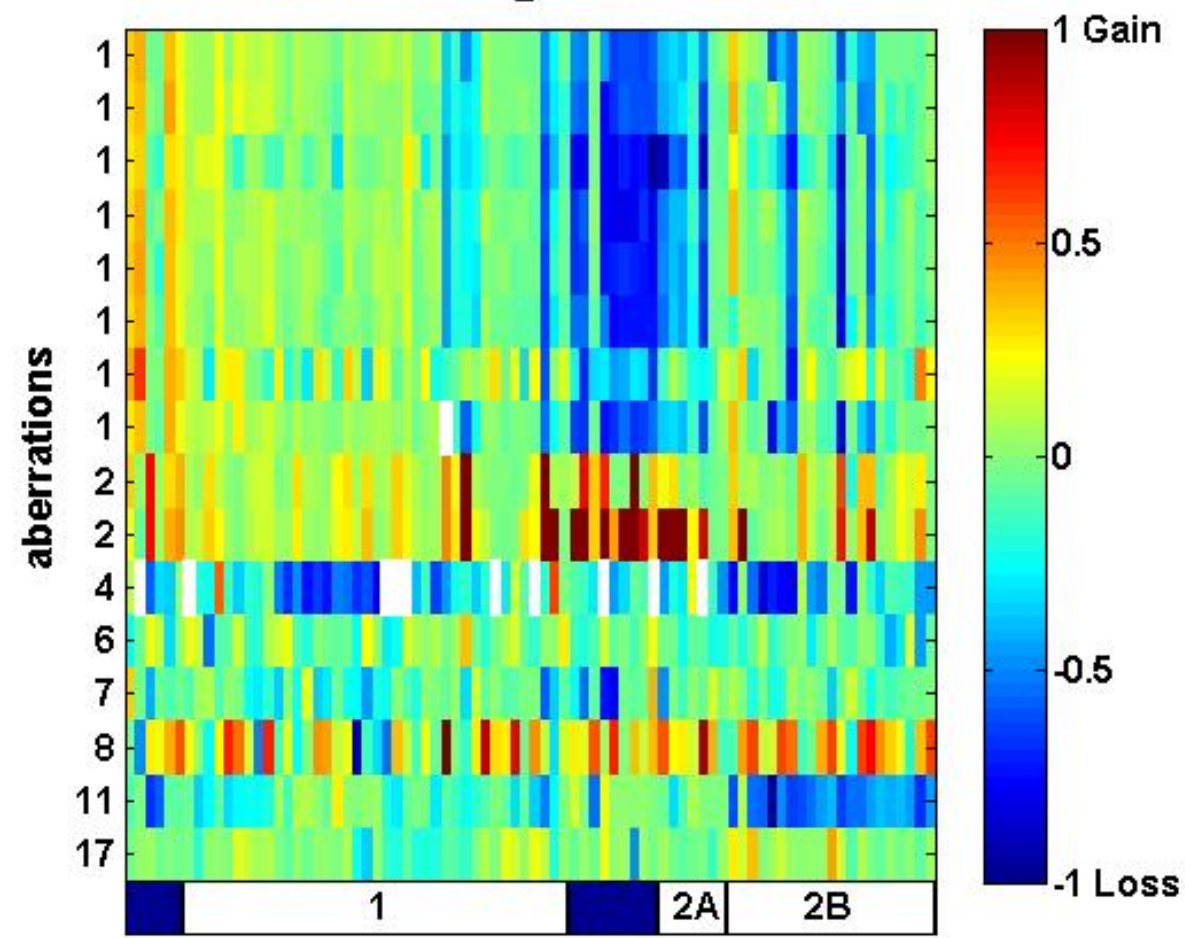